\documentclass[
preprint,   
 amsmath,amssymb,
 aps,
]{revtex4-1}
\usepackage{graphicx}
\usepackage{dcolumn}
\usepackage{bm}
\usepackage{graphics}
\usepackage{nicefrac, xfrac}

\usepackage{epstopdf}

\usepackage[inline]{enumitem}
\usepackage{tikz}
\usetikzlibrary{shapes.arrows}
\usetikzlibrary{decorations.pathmorphing}
\usetikzlibrary{decorations.markings}
\usetikzlibrary{decorations.pathmorphing}
\usetikzlibrary{decorations.pathreplacing}
\usetikzlibrary{math}
\usepgfmodule{oo}

\begin{document}

\title{ Simultaneous Multiphoton-Multiatom Processes in Atomic Gases under Laser Fields } 

\author{Yongle Yu }
\email{yongle.yu@wipm.ac.cn}

\affiliation{ State Key Laboratory of Magnetic Resonance and  \\ Atomic and 
Molecular Physics,
  Wuhan Institute of Physics \\ and Mathematics, Chinese Academy of Science,
 West No. 30 Xiao Hong Shan, Wuchang, Wuhan, 430071, China \vspace{2em}}

\begin{abstract}

We investigate simultaneous multiphoton-multiatom 
processes in atomic gases exposed to laser fields
under specific frequency conditions, 
where multiple atoms are simultaneously excited through
the absorption of  one laser photon each.  
These processes represent natural high-order 
quantum electrodynamics (QED) effects 
that occur independently of inter-atomic interactions.
A characteristic length scale emerges, governing 
the physical range over which these phenomena manifest.
 We propose experiments to 
demonstrate the fundamental aspects of these 
collective QED processes.

\end{abstract}
\maketitle

Quantum electrodynamics (QED) serves as the cornerstone for 
understanding the interaction between matter and light, providing 
profound insights into atomic and subatomic processes. When atoms 
are exposed to coherent electromagnetic fields, such as those 
produced by lasers, a wide range of phenomena emerges, spanning 
from single-photon absorption to intricate multiphoton transitions. 
Among these, multiphoton processes \cite{GoeppertMayer1931,twophoton,
 hernandez2004three, zheng2013frequency, chen2017giant}—where a single atom  
simultaneously absorbs multiple photons—represent high-order 
QED effects that have been extensively studied, leading to significant 
applications in nonlinear optics, laser physics, precision spectroscopy
and biological imaging.
However, relatively less attention has been directed toward another class 
of high-order QED processes: those involving the simultaneous transitions 
of multiple atoms, each absorbing a laser photon with a frequency that deviates significantly
 from its respective atomic transition frequency.
 These processes, characterized by 
joint atomic behavior, occur naturally within the framework of 
QED and represent an intriguing area for exploration in the 
interaction between light and matter. In this paper, we investigate some fundamental aspects
 of these multiphoton-multiatom (MPMA) processes. 
Notably, we highlight that such processes do not require mediation 
by interatomic interactions. Furthermore, we demonstrate that the 
transition rates of these processes can be substantially enhanced 
in the presence of a large number of atoms, which could lead to
new possibilities for studying ultra-weak atomic phenomena. We 
also propose experiments to  directly test several 
 theoretical findings presented in this work.

To provide a general illustration of multiphoton-multiatom  processes, 
we begin with an analysis of a two-photon-two-atom process.
A number of theoretical studies have addressed two-photon-two-atom processes, 
including those in \cite{rios1980lineshape, andrews1983cooperative, 
 Nayfeh1984, Kim1998, Muthukrishnan2004, Zheng2013}.
  Notably, two pioneering experimental 
works \cite{WhiteAtomgasTlBa, NaAtoms} observed such processes and 
offer comparative insights into their analysis. Furthermore, an analogous 
two-photon-two-molecule process has also been reported \cite{twoMolecules}.
 Consider a system of two non-interacting atoms, with one atom 
being of species $A$ and the other of species $B$, exposed 
to a homogeneous 
laser field (see Fig.~\ref{fig_twoAtom}). Each atom is assumed to be a two-level system, 
and the transition between 
its levels occurs via a dipole interaction with the electromagnetic field.
Denote the ground state of the $A$-species atom by $|g_a\rangle$
with an energy $\varepsilon^a_g$ and its excited state by 
 $|e_a\rangle$ with an energy $\varepsilon^a_e$. For the 
$B$-species atom, 
 the ground state $|g_{b}\rangle $
 has an energy of $\varepsilon^b_g$, and the excited state  $|e_{b}\rangle$
has an energy of $\varepsilon^b_e$.
The  angular transition 
frequency of the $A$-species atom is  $\omega_a= (\varepsilon^a_e - \varepsilon^a_g)/{\hbar}$,
 where $\hbar$ is Planck’s constant, while the  angular transition 
 frequency of the $B$-species atom 
is  $\omega_b= (\varepsilon^b_e - \varepsilon^b_g)/{\hbar}$. 
Although we introduce two distinct atomic species for formal simplicity, the 
multiphoton-multiatom process can also be realized using a gas composed of a 
single atomic species. In such cases, two different excitation states of 
the same species, each with a distinct transition frequency, can be 
selectively targeted, as demonstrated in the experiment reported in \cite{NaAtoms}.

The angular frequency of the laser is denoted by $\Omega_\mathfrak{L}$, set to be
 $ \Omega_\mathfrak{L} = 
 (\omega_a + \omega_b)/2$. It is 
 assumed that $\omega_a$ and $\omega_b$ differ significantly. 
 Under  this configuration, one can easily note that 
 if both atoms are initially in their ground states, 
 neither atom can individually absorb a single photon due to the mismatch 
 between $\Omega_\mathfrak{L}$ and the respective transition frequency
  $\omega_a$ or $\omega_b$. However, a joint excitation of the two atoms 
  can occur through simultaneous absorption of one photon by each atom, 
  satisfying 
  energy conservation in the two-atom process.

 \tikzmath{ \bx=-1.2; \bz= 0.4; \bsz=0.5; \bsf=6;    
\wsf=\bsf+5; 
\wint= 0.12;  
\wx= -3.8; \wy=0.8; \wz= 0.7; 
\wsz=1.5;
\asz=0.07;
\rEx= 0.1; \Esz=0.7; \rEy= -1.4;  \rEz=0.2;\rint=2.2; 
\bEx= -0.1;  \bEy= 1.0; \bEz= -0.1;  \bint= 0.8; \brEyg=0.3; 
\Elabsz= 0.85; 
\atomrEgapx=1.2;\atombEgapx= 1.0; \atombExn=-0.18; 
\rbmix= 0.65; \rbmixn= 1-\rbmix;
\atomsize= 0.6;  
\pad= 0.05;
\seglen= 8pt; \amp=1.pt; 
\padint= 1.3pt;
 }
\definecolor{algeagreen}{RGB}{100, 233, 134}
\definecolor{redgold}{RGB}{235, 84, 6}

\pgfooclass{stamp}{

 \attribute text;
  \attribute rotation angle = 20;
   \attribute colors;
  \method stamp(#1) {
    \pgfooset{text}{#1} 
   }
   \method setcolor(#1) {
    \pgfooset{colors}{#1} 
   }

  \method apply(#1,#2,#3) {

    \node [rotate=\pgfoovalueof{rotation angle},font=\huge, color=#3] at(#1,#2)
      {\pgfoovalueof{text}};
  }
}

\pgfooclass{photonline}{
   \method apply(#1,#2,#3,#4,#5,#6) {
   \draw[decorate,decoration={snake, segment length= \seglen, amplitude=\amp},color=#6](#1, #2)  -- +(#3,0);
   \draw[-stealth,color=#6](#1+#4,#2)  -- +(#5,0);
  
  }

 \method apply_dashed(#1,#2,#3,#4,#5,#6) {
 \draw[decorate,dash pattern={on \padint off \padint on \padint off \padint on \padint},
 decoration={snake, segment length= \seglen, amplitude=\amp},color=#6](#1, #2)  -- +(#3,0);
  \draw[-stealth,color=#6](#1+#4,#2)  -- +(#5,0);
  
  }
 }

\pgfooclass{gravitonline}{
   \method apply(#1,#2,#3,#4,#5,#6) {
   \draw[decorate,decoration={coil, segment length= \seglen, amplitude=\amp},color=#6](#1, #2)  -- +(#3,0);
   \draw[-stealth,color=#6](#1+#4,#2)  -- +(#5,0);
  
  }

 \method graviton_dashed(#1,#2,#3,#4,#5,#6) {
 \draw[decorate,dash pattern={on \padint off \padint on \padint off \padint on \padint},
 decoration={coil, segment length= \seglen, amplitude=\amp},color=#6](#1, #2)  -- +(#3,0);
  \draw[-stealth,color=#6](#1+#4,#2)  -- +(#5,0);
  
  }
 }

\pgfoonew \mystamp=new stamp()
\pgfoonew \myphotonline=new photonline()
\pgfoonew \mygravitonline=new gravitonline()
\begin{tikzpicture}
\mystamp.stamp(use this)
\mystamp.setcolor(red)
\end{tikzpicture}

\tikzmath{ \arcR=40; \arcAng=5; \arcH=1.8;  \opacity=0.7; \ballR=0.1cm; 
   \centrX=4.4; \atomShift=0.9; \centrY=-0.9; \bEx=1; \bEy=0.5;\rEx= -1.0; \Elabsz= 0.85;
   \rEy= 0.5; \rint= 1.2;\bint=0.8; \bsz=0.5; \bsf=6; \pad=0.05; \seglen= 8pt; \amp=1.pt;\Esz=0.7; 
\padint= 1.3pt; }
\definecolor{LaserPink}{RGB}{241,184,186}
\definecolor{Laserblue_detune}{RGB}{20, 163, 199} 
\definecolor{LaseryellowGreen}{RGB}{226, 245, 22} 
\definecolor{atomgray}{RGB}{205,205,205}
\definecolor{arXivgray}{RGB}{218, 219, 221}
\definecolor{left} {HTML}{001528}
\colorlet{rcol}{green!20!blue!80}
\colorlet{bcol}{magenta!60!}
\begin{figure}
\begin{tikzpicture}[scale=1.0]
\draw[draw=none, fill= algeagreen, opacity= \opacity,rotate= 0] (8,0)arc[start angle= 270+\arcAng, end angle= 270-\arcAng,radius=\arcR] -- +(0,-\arcH)
   arc[start angle= 90+\arcAng, end angle= 90-\arcAng,radius=\arcR]-- cycle;
  \node[font = {\tiny}] (c) at (7., -1.5) {${\Omega_\mathfrak{L}=\frac{\omega_a+\omega_b}{2}}$ }; 
 
  \node [circle, inner color= rcol, outer color= atomgray, minimum size=\ballR, opacity= \opacity, font={\tiny}]at(\centrX-\atomShift,\centrY) {$A$};  
  \node [circle, inner color= bcol, outer color= atomgray, minimum size=\ballR, opacity= \opacity, font={\tiny}]at(\centrX+\atomShift,\centrY) {$B$};  
 
  \draw[thick, rcol](\rEx+\bsf, \rEy)--+(\Esz, 0);
\draw[thick, rcol](\rEx+\bsf, \rEy+\rint)--+(\Esz, 0);
\draw[thick,rcol](\rEx +\Esz/2+\bsf, \rEy) circle (0.04);
\draw[dash pattern={on \padint off \padint on \padint off \padint on \padint}](\centrX-\atomShift+2*\pad,\centrY+2*\pad)--(\rEx+\bsf - \pad, \rEy);
\draw[dash pattern={on \padint off \padint on \padint off \padint on \padint}](\centrX-\atomShift+2*\pad,\centrY+3*\pad)--(\rEx+\bsf - \pad, \rEy+\rint);
\draw[dashed, rcol, -stealth](\rEx+\Esz/2+\bsf, \rEy)--+(0, \rint);
\node[font = {\tiny},color=rcol] (c) at (\rEx+\Esz/2+\bsf+ 4*\pad ,\rEy+\rint/2) {$\omega_a$ };
  
  \draw[thick, bcol](\bEx+\bsf, \bEy)--+(\Esz, 0);
\draw[thick, bcol](\bEx+\bsf, \bEy+\bint)--+(\Esz, 0);
\draw[thick,bcol](\bEx +\Esz/2+\bsf, \bEy) circle (0.04);
\draw[dash pattern={on \padint off \padint on \padint off \padint on \padint}](\centrX+\atomShift+2*\pad,\centrY+2*\pad)--(\bEx+\bsf - \pad, \bEy);
\draw[dash pattern={on \padint off \padint on \padint off \padint on \padint}](\centrX+\atomShift+2*\pad,\centrY+3*\pad)--(\bEx+\bsf - \pad, \bEy+\bint);
\draw[dashed, bcol, -stealth](\bEx+\Esz/2+\bsf, \bEy)-- +(0, \bint);
\node[font = {\tiny},color=bcol] (c) at (\bEx+\Esz/2+\bsf+ 4*\pad ,\bEy+\bint/2) {$\omega_b$ };

   

\end{tikzpicture}
\caption{A system of two non-interacting atoms in a laser field.}
\label{fig_twoAtom}
\end{figure}
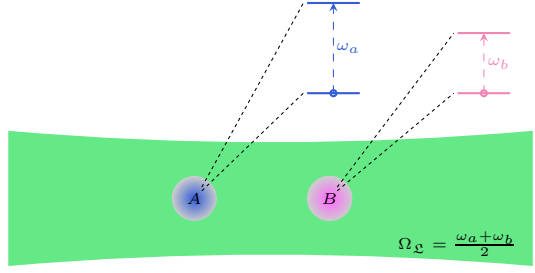

 The Hamiltonian of 
 this atomic system  and the laser field  can be written as:
 \begin{equation}
{\hat H} =  {\hat H_a} + {\hat H_b} + \hbar \Omega_\mathfrak{L} {\hat a^\dagger}{\hat a}  +
         \hat{\mathbf d}_a \cdot {\mathbf {\hat E}} + \hat{\mathbf d}_b\cdot {\mathbf {\hat E}}.
 \end{equation}
Here, $\hat H_a = \varepsilon^a_g |g_a \rangle \langle g_a| + \varepsilon^a_e |e_a\rangle \langle e_a|$, and 
$\hat H_{b} = \varepsilon^b_g |g_{b} \rangle \langle g_{b}| + \varepsilon^b_e |e_b\rangle \langle e_{b}|
 $, represent the atomic Hamiltonians of the $A$-species atom and  the $B$-species atom respectively.
  $\hat{a} (\hat{a}^\dagger)$ is the  annihilation (creation) operator 
 of the laser photon.
       The operators $\hat{\mathbf d}_a $ and $\hat{\mathbf d}_{b}$ are
 the dipole moments of the $A$-species atom and the $B$-species atom, respectively. ${\mathbf{\hat{E}}}$ is
 the quantum operator of the laser's electric field. The dipole moment operators can be written as:
 \begin{equation}
\hat{\mathbf d}_k= \langle e_k |e {\mathbf{\hat{r}}}_k | g_k \rangle |g_k \rangle \langle e_k|
 +  \langle g_k |e \mathbf{\hat{ r}}_k | e_k \rangle |e_k \rangle \langle g_k| \ (k=a, b).
\end{equation}
Here, $e$ is the electric charge of the electron, and $ {\mathbf {\hat{r}}}_{a} $ and 
 ${\mathbf {\hat{r}}}_{b}$ are the position operators of the electron in the $A$-species atom
 and the $B$-species atom, respectively.
 $\mathbf{\hat{E}}$ can be approximately written in the form:
 \begin{equation}
 {\mathbf {\hat E}}= i (\frac {\hbar \Omega_\mathfrak{L}}{2 \epsilon_0 V})^{1/2} {\mathbf e}( {\hat a}e^{-i\Omega_\mathfrak{L} t}-{\hat a^\dagger}e^{i\Omega_\mathfrak{L} t}),
\end{equation}
 where $V$ is the volume of the laser field, $\epsilon_0$ is the 
permittivity of free space, and $\mathbf{e}$ is the polarization 
vector.

\begin{figure}
\begin{tikzpicture}[scale=1.0] 

\draw[thick, rcol](\rEx +\levelshift, \rEy)-- +(\Esz, 0);
\node[font = {\tiny},color=rcol] (c) at (\rEx+\Esz*\Elabsz+ \levelshift,\rEy+3.5*\pad) {${\varepsilon^a_g}$ };
\draw[thick, rcol](\rEx+\levelshift, \rEy+\rint)--+(\Esz, 0); 
\node[font = {\tiny},color=rcol] (c) at (\rEx+\Esz*\Elabsz+\levelshift,\rEy+\rint+3.5*\pad) {${\varepsilon^a_e}$ };
\draw[thick,rcol](\rEx +\Esz/2+\levelshift, \rEy) circle (0.04);

\draw[dashed, rcol, -stealth](\rEx+\Esz/2+\levelshift, \rEy)-- +(0, \rint);

\draw[thick, bcol](\bEx+\levelshift, \bEy)-- +(\Esz, 0);
\node[font = {\tiny},color=bcol] (c) at (\bEx+\Esz*\Elabsz+\levelshift,\bEy+3.5*\pad) {${\varepsilon^b_g}$ };
\draw[thick, bcol](\bEx+\levelshift, \bEy+\bint)-- +(\Esz, 0);
\node[font = {\tiny},color=bcol] (c) at (\bEx+\Esz*\Elabsz+\levelshift,\bEy+\bint+3.5*\pad) {${\varepsilon^b_e}$ };
\draw[thick,bcol](\bEx +\Esz/2+\levelshift, \bEy) circle (0.04);


\myphotonline.apply(\wx, 0, \wsz, -2*\wint+\wsz, \asz, algeagreen)
\node[font = {\tiny}, color= algeagreen] (c) at (\wx+ 0.5*\wsz, \wyinz) { ...};
\myphotonline.apply(\wx, 2*\wyinz, \wsz, -2*\wint+\wsz, \asz,  algeagreen)
\myphotonline.apply(\wx, 3*\wyinz, \wsz, -2*\wint+\wsz, \asz,  algeagreen)
\myphotonline.apply_dashed(\wx, 4*\wyinz, \wsz, -2*\wint+\wsz, \asz,  algeagreen)
\draw[thick](\wx-0.1, -\wyinz)  -- +(0, 6*\wyinz);
\draw[thick](\wx+ \wsz + 0.1, 5*\wyinz)  -- +(0.1, -3*\wyinz) -- +(0, -6*\wyinz) ;
\node[font = {\tiny}] (c) at (\wx+ \wsz+ 0.25, -\wyinz) {n};
\myphotonline.apply(\wx, 8*\wyinz, \wsz, -2*\wint+\wsz, \asz,  algeagreen)
\draw[dashed, dash pattern={on \padint off \padint on \padint off \padint on \padint}, -stealth]
(\wx + 0.5* \wsz , 4.8*\wyinz )-- +(0, 2.4*\wyinz);

\draw[thick, rcol](\rEx + \xpanel+\levelshift, \rEy)-- +(\Esz, 0);
\node[font = {\tiny},color=rcol] (c) at (\rEx+\Esz*\Elabsz+ \xpanel+\levelshift,\rEy+3.5*\pad) {${\varepsilon^a_g}$ };
\draw[thick, rcol](\rEx+ \xpanel+\levelshift, \rEy+\rint)--+(\Esz, 0); 
\node[font = {\tiny},color=rcol] (c) at (\rEx+\Esz*\Elabsz+ \xpanel+\levelshift,\rEy+\rint+3.5*\pad) {${\varepsilon^a_e}$ };
\draw[thick,rcol](\rEx +\Esz/2+ \xpanel+\levelshift, \rEy) circle (0.04);


\draw[thick, bcol](\bEx+ \xpanel+\levelshift, \bEy)-- +(\Esz, 0);
\node[font = {\tiny},color=bcol] (c) at (\bEx+\Esz*\Elabsz+ \xpanel+\levelshift,\bEy+3.5*\pad) {${\varepsilon^b_g}$ };
\draw[thick, bcol](\bEx+ \xpanel+\levelshift, \bEy+\bint)-- +(\Esz, 0);
\node[font = {\tiny},color=bcol] (c) at (\bEx+\Esz*\Elabsz+ \xpanel+\levelshift,\bEy+\bint+3.5*\pad) {${\varepsilon^b_e}$ };
\draw[thick,bcol](\bEx +\Esz/2+ \xpanel+\levelshift, \bEy) circle (0.04);

\draw[dashed, bcol, -stealth](\bEx+\Esz/2+ \xpanel+\levelshift, \bEy)-- +(0,\bint); 

\myphotonline.apply(\wx+ \xpanel, 0, \wsz, -2*\wint+\wsz, \asz, algeagreen)
\node[font = {\tiny}, color= algeagreen] (c) at (\wx+ 0.5*\wsz+ \xpanel , \wyinz) { ...};
\myphotonline.apply(\wx+ \xpanel, 2*\wyinz, \wsz, -2*\wint+\wsz, \asz,  algeagreen)
\myphotonline.apply(\wx+ \xpanel, 3*\wyinz, \wsz, -2*\wint+\wsz, \asz,  algeagreen)
\myphotonline.apply_dashed(\wx+ \xpanel, 4*\wyinz, \wsz, -2*\wint+\wsz, \asz,  algeagreen)
\draw[thick](\wx-0.1+ \xpanel, -\wyinz)  -- +(0, 6*\wyinz);
\draw[thick](\wx+ \wsz + 0.1+ \xpanel, 5*\wyinz)  -- +(0.1, -3*\wyinz) -- +(0, -6*\wyinz) ;
\myphotonline.apply(\wx+ \xpanel, 8*\wyinz, \wsz, -2*\wint+\wsz, \asz,  algeagreen)
\draw[dashed, dash pattern={on \padint off \padint on \padint off \padint on \padint}, -stealth]
(\wx + 0.5* \wsz+ \xpanel , 4.8*\wyinz )-- +(0, 2.4*\wyinz);

 
 \node[single arrow,draw,  dash pattern={on \padint off \padint on \padint off \padint on \padint}, rotate=90,
      minimum width = 16pt, single arrow head extend= 5pt,
      minimum height=5.0mm] at (\wx + 1.2*\Esz+ 1.*\wsz + 0.0*\xpanel, 0.65* \ypanel  ) {};
      
 \node[single arrow,draw,  dash pattern={on \padint off \padint on \padint off \padint on \padint}, rotate=90,
      minimum width = 16pt, single arrow head extend= 5pt,
      minimum height=5.0mm] at (\wx + 1.2*\Esz+ 1.*\wsz + 1.0*\xpanel, 0.65* \ypanel  ) {};
      
  \node[single arrow,draw,  dash pattern={on \padint off \padint on \padint off \padint on \padint}, rotate= 45,
      minimum width = 16pt, single arrow head extend= 5pt,
      minimum height=5.0mm] at (\wx + 1.2*\Esz+ 1.*\wsz+0.3*\xpanel, 1.75* \ypanel  ) {};

  \node[single arrow,draw,  dash pattern={on \padint off \padint on \padint off \padint on \padint}, rotate= 135,
      minimum width = 16pt, single arrow head extend= 5pt,
      minimum height=5.0mm] at (\wx + 1.2*\Esz+ 1.*\wsz + 0.7* \xpanel , 1.75* \ypanel  ) {};
      
  
 \draw[draw=none, fill=arXivgray, opacity= 0.2,rotate= 0] (\wx-0.5, \rEy+\ypanel-0.5)-- +(0, 1.3*\rint+0.5 ) --
 +(0.5 + 2.5*\Esz+ 2*\wsz,  1.3*\rint+0.5)-- +(0.5+ 2.5*\Esz+ 2*\wsz, 0 )--cycle;

\draw[draw=none, fill= arXivgray, opacity= 0.2,rotate= 0] (\wx-0.5+\xpanel, \rEy+\ypanel-0.5)-- +(0, 1.3*\rint+0.5 ) --
 +(0.5 + 2.5*\Esz+ 2*\wsz,  1.3*\rint+0.5)-- +(0.5+ 2.5*\Esz+ 2*\wsz, 0 )--cycle;
  
 \draw[thick, rcol](\rEx, \rEy+\ypanel)--+(\Esz, 0);
\draw[thick, rcol](\rEx, \rEy+\rint+\ypanel)--+(\Esz, 0);
\draw[thick,rcol](\rEx +\Esz/2, \rEy+\ypanel+\rint) circle (0.04);

 \draw[thick, bcol](\bEx, \bEy+\ypanel)--+(\Esz, 0);
\draw[thick, bcol](\bEx, \bEy+\bint+\ypanel)-- +(\Esz, 0);
\draw[thick,bcol](\bEx +\Esz/2, \bEy+\ypanel) circle (0.04);

\draw[dashed, bcol, -stealth](\bEx+\Esz/2, \bEy+\ypanel)--+(0,\bint);

\draw[thick](\wx-0.1, -\wyinz+\ypanel)  -- +(0, 6*\wyinz);
\draw[thick](\wx+ \wsz + 0.1, 5*\wyinz+\ypanel)  -- +(0.1, -3*\wyinz) -- +(0, -6*\wyinz) ;

\draw[dashed, dash pattern={on \padint off \padint on \padint off \padint on \padint}, -stealth](\wx + 0.5* \wsz , 4.8*\wyinz+\ypanel )--
+(0, 2.4*\wyinz);
\myphotonline.apply(\wx, \ypanel, \wsz, -2*\wint+\wsz, \asz,  algeagreen)
\node[font = {\tiny},color= algeagreen] (c) at (\wx+ 0.5*\wsz , 1.5*\wyinz+\ypanel) { ...};
\myphotonline.apply(\wx,  3*\wyinz+\ypanel, \wsz, -2*\wint+\wsz, \asz,  algeagreen)
\myphotonline.apply_dashed(\wx,  4*\wyinz+\ypanel, \wsz, -2*\wint+\wsz, \asz,  algeagreen)
\myphotonline.apply(\wx,  8*\wyinz+\ypanel, \wsz, -2*\wint+\wsz, \asz,  algeagreen)
 \draw[thick, rcol](\rEx + 0.5*\xpanel, \rEy+2*\ypanel)--  +(\Esz,0);
\draw[thick, rcol](\rEx +0.5*\xpanel, \rEy+\rint+2*\ypanel)-- +(\Esz,0);
\draw[thick,rcol](\rEx +\Esz/2 +0.5*\xpanel, \rEy+\rint+ 2*\ypanel) circle (0.04);

  \draw[thick, bcol](\bEx +0.5*\xpanel, \bEy+2*\ypanel)-- +(\Esz,0);
\draw[thick, bcol](\bEx+0.5*\xpanel, \bEy+\bint+2*\ypanel)--  +(\Esz,0);

\draw[thick,bcol](\bEx +\Esz/2+0.5*\xpanel, \bEy+\bint+2*\ypanel) circle (0.04);

\draw[thick](\wx-0.1+0.5*\xpanel, -\wyinz+2*\ypanel)  -- +(0, 6*\wyinz);
\draw[thick](\wx+ \wsz + 0.1+0.5*\xpanel, 5*\wyinz+2*\ypanel)  -- +(0.1, -3*\wyinz) -- +(0, -6*\wyinz) ;

\myphotonline.apply(\wx+0.5*\xpanel, 2*\ypanel, \wsz, -2*\wint+\wsz, \asz,  algeagreen)
\node[font = {\tiny},color= algeagreen] (c) at (\wx+0.5*\xpanel+ 0.5*\wsz , 2*\wyinz+2*\ypanel) { ...};
\myphotonline.apply(\wx+0.5*\xpanel,  4*\wyinz+2*\ypanel, \wsz, -2*\wint+\wsz, \asz,  algeagreen)


    
  
 \draw[thick, rcol](\rEx+\xpanel, \rEy+\ypanel)--+(\Esz, 0);
\draw[thick, rcol](\rEx+\xpanel, \rEy+\rint+\ypanel)--+(\Esz, 0);
\draw[thick,rcol](\rEx +\Esz/2+\xpanel, \rEy+\ypanel) circle (0.04);
\draw[dashed, rcol, -stealth](\rEx+\Esz/2+\xpanel, \rEy+\ypanel)--+(0,\rint);
   
 \draw[thick, bcol](\bEx+\xpanel, \bEy+\ypanel)--+(\Esz, 0);
\draw[thick, bcol](\bEx+\xpanel, \bEy+\bint+\ypanel)-- +(\Esz, 0);
\draw[thick,bcol](\bEx +\Esz/2+\xpanel, \bEy+\ypanel+\bint) circle (0.04);


\draw[thick](\wx-0.1+\xpanel, -\wyinz+\ypanel)  -- +(0, 6*\wyinz);
\draw[thick](\wx+ \wsz + 0.1+\xpanel, 5*\wyinz+\ypanel)  -- +(0.1, -3*\wyinz) -- +(0, -6*\wyinz) ;

\draw[dashed, dash pattern={on \padint off \padint on \padint off \padint on \padint}, -stealth]
(\wx + 0.5* \wsz+\xpanel , 4.8*\wyinz+\ypanel )--
+(0, 2.4*\wyinz);

\draw[dashed, dash pattern={on \padint off \padint on \padint off \padint on \padint}, -stealth](\wx + 0.5* \wsz , 4.8*\wyinz+\ypanel )--
+(0, 2.4*\wyinz);
\myphotonline.apply(\wx+\xpanel, \ypanel, \wsz, -2*\wint+\wsz, \asz,  algeagreen)
\node[font = {\tiny},color= algeagreen] (c) at (\wx+\xpanel+ 0.5*\wsz , 1.5*\wyinz+\ypanel) { ...};
\myphotonline.apply(\wx+\xpanel,  3*\wyinz+\ypanel, \wsz, -2*\wint+\wsz, \asz,  algeagreen)
\myphotonline.apply_dashed(\wx+\xpanel,  4*\wyinz+\ypanel, \wsz, -2*\wint+\wsz, \asz,  algeagreen)
\myphotonline.apply(\wx+\xpanel,  8*\wyinz+\ypanel, \wsz, -2*\wint+\wsz, \asz,  algeagreen)


  

   

\end{tikzpicture}  
\caption{The excitation pathways for the two-photon-two-atom process, with quantum states in 
shaded boxes representing virtual intermediate states involved in the transitions.
}
\label{pathways}
\end{figure}
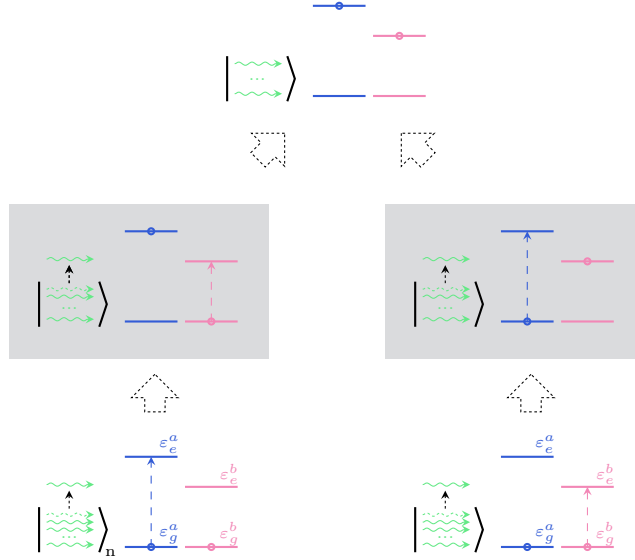

Assume the initial state of the whole system, including 
the laser field, is $|\Psi_i\rangle = |g_a\rangle |g_{b}\rangle 
| \scalebox{0.9}{$N_\gamma$} \rangle$  
($| \scalebox{0.9}{$N_\gamma$} \rangle $ represents the quantum state of the
laser field with $\scalebox{0.9}{$N_\gamma$}$ laser photons). At time $t$, the probability of the system being 
in the excited state $|\Psi_f\rangle = |e_a\rangle |e_b\rangle |\scalebox{0.9}{$N_\gamma-2$}\rangle$ is
\begin{equation}
P_{ex} = |\langle \Psi_f | e^{-i\int_0^t \hat{H} dt^\prime/\hbar} | \Psi_i \rangle |^2,
\end{equation}
which will not be zero in principle. In the interaction picture, the third term of 
the perturbation expansion of $e^{-i\int_0^t \hat{H} dt^\prime/\hbar}$ 
can couple $|\Psi_i\rangle$ to $|\Psi_f\rangle$ 
through  virtual intermediate states. 
Defining the following
intermediate states: $|\Psi_{a}\rangle=
 |e_a\rangle |g_{b}\rangle  
|\scalebox{0.9}{$N_\gamma-1$} \rangle$,  $|\Psi_{b}\rangle = |g_a\rangle |e_{b}\rangle  
|\scalebox{0.9}{$N_\gamma-1$}\rangle$. The two-photon excitations
 can occur through two virtual quantum transition paths,
which are $|\Psi_i\rangle \rightarrow |\Psi_{a}\rangle  \rightarrow  |\Psi_f\rangle$ and  $|\Psi_i\rangle \rightarrow |\Psi_{b}\rangle\rightarrow |\Psi_f\rangle$ (see Fig.~\ref{pathways}).
These transitions correspond to a simultaneous 
two-photon-two-atom process in QED, where 
the energy of the system is not conserved for the virtual intermediate states.

\tikzmath{ \arcR=40; \arcAng=5; \arcH=1.8;  \opacity=0.7; \ballR=0.1cm; 
   \centrX=4.4; \atomShift=0.9; \centrY=-0.9; \bEx=1;\rEx= -0.86;  \Elabsz= 0.85;
   \rEy= -0.03; \bEy=\rEy; \rint= 1.2;\bint=0.8; \bsz=0.5;  \pad=0.05; \seglen= 5pt; \amp=0.6pt;\Esz=0.7; 
\padint= 1.pt; \wsz=0.6; \bEx= \rEx + \Esz+ 0.1; \bsf=0; \wint=0.02; \wx=-1.9; \levelshift=-0.0; \wyinz=0.1;\ypanel=3.0;   \xpanel=5.0;}

 The transition rate for the simultaneous two-atom excitation 
 can be calculated using the usual perturbation method.
 After applying the rotating wave approximation, it can be written in the following form: 
\begin{equation}
W_{2p2m}= \frac{2\pi}{\hbar}
\left|\sum_{\kappa}\frac{\langle \Psi_f | H_{int}| \Psi_{\kappa}\rangle   \langle \Psi_{\kappa}  | H_{int}|\Psi_i\rangle}
{ \varepsilon^a_g + \varepsilon^b_g + \scalebox{0.9}{$N_\gamma$}\hbar \Omega_\mathfrak{L} - E_{\Psi_\kappa}} \right|^2\rho(E_f)|_{E_f= \varepsilon^a_g +\varepsilon^b_g +2 \hbar \Omega_\mathfrak{L}}.
\end{equation} 
Here,  $H_{int}=  \hat{\mathbf d}_a \cdot {\mathbf {\hat E}} + \hat{\mathbf d}_{b}\cdot {\mathbf {\hat E}}$,
is the coupling between the atoms and laser light. The state $\Psi_{\kappa}= \Psi_{a} $ or $\Psi_{b}$, is
 one of the intermediate states with energy
  $E_{\Psi_\kappa} = \varepsilon_\kappa + (\scalebox{0.9}{$N_\gamma-1$})\hbar \Omega_\mathfrak{L} \,(\kappa= a,b)$.
The factor $\rho(E_f)$ corresponds to the level density of the two-atom 
system and relates to the level density of each atom \cite{densityStates}.

The transition rate for the process can be computed and written in the form:  
\begin{equation}
W_{2p2m} \approx \frac{\hbar}{2^7 \pi^5} {(\Omega_\mathfrak{L}/\omega_a)}{(\Omega_\mathfrak{L}/\omega_b)}{\gamma_a \gamma_b \Omega_\mathfrak{L}^2} n_{\lambda^3}^2\left| \frac{1}{\Omega_\mathfrak{L} - \omega_a}+ \frac{1}{\Omega_\mathfrak{L} - \omega_b}\right|^2 \rho(E_f).
\label{eq:W2}
 \end{equation} 
Here, $\gamma_a=  {4 \alpha_e  \omega_a^3} |\langle e_a| {\mathbf{\hat{r}_a}}| g_a \rangle|^2/ {3 c^2}$ and
$\gamma_b= {4 \alpha_e   \omega_b^3} |\langle e_b| {\mathbf{\hat{r}_b}}| g_b \rangle|^2/ {3 c^2}$, where 
$\alpha_e = {e^2}/{ 4\pi \hbar c \epsilon_0} \approx 1/137 $ is the fine-structure constant, and $c$ is the speed of
light.
Additionally, $n_{\lambda^3}$ represents the number of laser photons within a volume of
$\lambda^3 = (2\pi c/\Omega_\mathfrak{L})^3$, defined as $n_{\lambda^3} =  \scalebox{1.0}{$N_\gamma$}  \lambda^3 /V$,
and it is assumed that  $N_\gamma \gg 1 $.

In this plain perturbation calculation, one can note that
$W_{2p2m}$ vanishes exactly since the two terms within the modulus on the right 
side of the Eq.~(\ref{eq:W2}) cancel each other at $\Omega_\mathfrak{L}= (\omega_a + \omega_b)/2$.
This result can be viewed as a  quantum interference effect 
between the two excitation pathways: $|\Psi_i\rangle \rightarrow |\Psi_{a}\rangle  \rightarrow  |\Psi_f\rangle$ and  $|\Psi_i\rangle \rightarrow |\Psi_{b}\rangle\rightarrow |\Psi_f\rangle$.

However, this cancellation does not imply the complete absence of joint two-atom 
excitation processes. Instead, two considerations suggest that quantum interference
 introduces a suppression factor for the transition rate rather than exact vanishing:
 
i) {\it Higher-order QED contributions}: when higher-order QED processes are 
taken into account, $W_{2p2m}$ no longer vanishes.  In particular,
 the introduction of a finite 
width for each atom's excited level modifies
the interference terms. By incorporating an imaginary component into $\omega_a$ and $\omega_b$ 
in Eq.~(\ref{eq:W2}), the terms within the modulus transform as follows:
 \begin{equation}
 \begin{split}
\left|\frac{1}{\Omega_\mathfrak{L} - \omega_a + i\Gamma/2\hbar}+ \frac{1}{\Omega_\mathfrak{L} - \omega_b+i\Gamma/2\hbar}\right|^2 & = \frac{\Gamma^2/\hbar^2}{((\Omega_\mathfrak{L}-\omega_b)^2 + \Gamma^2/4\hbar^2)^2} \\
& \approx \frac{\Gamma^2/\hbar^2}{(\Omega_\mathfrak{L}-\omega_b)^2} \frac{1}{(\Omega_\mathfrak{L}-\omega_b)^2},
\end{split}
\label{eq:wlifewidth} 
\end{equation}
where $\Gamma$ represents the (natural) width of the excited level, assumed identical for 
both atoms for simplicity.  In QED, this width arises from a series of repeated 
virtual processes, wherein an excited atom transitions to the 
ground state by emitting a virtual photon and reabsorbs it to return
to the excited state (see,{\it e.g.}, \cite{Tannoudji}). 
 In the context of Eq.~(\ref{eq:wlifewidth}),
the factor $\frac{\Gamma^2/\hbar^2}{(\Omega_\mathfrak{L} - \omega_b)^2}$,  typically orders of 
magnitude smaller than unity, can be interpreted as the suppression factor 
associated with the quantum interference.

ii) {\it Effect of detuning:} The formal cancellation strictly occurs only at the specific value
 $\Omega_\mathfrak{L}= (\omega_a + \omega_b)/2$. However, in practice, $\Omega_\mathfrak{L}$ 
 can be set to cover a range of frequencies, resulting in detuning. 
 Considering
$\Omega_\mathfrak{L} = (\omega_a + \omega_b)/2 + \Delta\Omega/2$, where $\Delta\Omega$ 
is a detuning parameter with a   magnitude smaller than or comparable to $\Gamma/\hbar$,
 the cancellation is no longer exact even in the plain perturbation treatment.  
In this case, the joint two-atom excitation involves 
additional processes, such as spontaneous photon emission.
For instance, two atoms may jointly transition to excited states by 
absorbing one laser photon each, followed by the 
$A$-species atom emitting a photon and returning to its ground state.
The emitted photon has a frequency $\omega_a + \Delta\Omega$ such 
that energy conservation 
for the entire process is satisfied. In this scenario, 
the final state of the system can be expressed as
$|g_a\rangle |e_b\rangle |N_\gamma - 2\rangle |\gamma^{sp}\rangle_{\Omega_\gamma = \omega_a + \Delta\Omega}$,
where $|\gamma^{sp}\rangle_{\Omega_\gamma}$ represents the quantum state
of the emitted photon at frequency $\Omega_\gamma$.
The transition rate of this joint two-atom process 
involving one-photon emission 
can be obtained similarly using perturbation theory. One is generally
interested in the emitted-photon exclusive transition rate, which is
obtained by summing  the transition rate over all 
possible directions of the emitted photon.
The exclusive rate 
near $\Delta\Omega \approx \Gamma/\hbar$ is comparable to the transition rate for exact 
joint excitation at $\Omega_\mathfrak{L}= (\omega_a + \omega_b)/2$ 
(with the finite-width effect included for exact joint excitation). 
In principle, joint two-atom processes involving emissions of two photons 
 are also possible,
 but a detailed investigation of these cases is not pursued here. 
The focus  instead is  to highlight other fundamental features of two-photon-two-atom 
processes and general MPMA processes, 
which remain independent of the specific details of 
the transition rate.

General MPMA processes can be constructed analogously to two-photon-two-atom processes.
  Consider a system of $m$ atoms ($m=3,4,...$)
 exposed to a laser field, where one atom is of  species $A$
  and the remaining $m-1$ atoms are of  species $B$.
 If the laser field frequency satisfies the condition 
 $ m\Omega_\mathfrak{L} = \omega_a + (m-1)\omega_b$, and the system 
 starts with all atoms in their ground states, then all atoms can be 
 simultaneously excited by absorbing  $m$ photons from the laser field.
In this $m$-th order QED process, the system progresses through 
a sequence of virtual intermediate states, with each atom becoming excited by 
absorbing one photon.  The transition rate for this process can be estimated using perturbation theory.
Similar to a two-photon-two-atom process, quantum interference among different 
excitation pathways arises, introducing a suppression factor in the transition rate \cite{quantum_inference}.
 Moreover, when
$\Omega_\mathfrak{L}$ deviates from the exact resonance frequency $\omega_a/m + (m-1)\omega_b/m$,
the excitation process in the 
$m$-atom system becomes coupled with spontaneous photon emission.

 An intriguing aspect of MPMA processes is their ability to 
 facilitate joint quantum behavior among multiple atoms without
 mediation of any interatomic interactions.  Similarly, another 
 remarkable joint quantum phenomenon that occurs without 
 mediated interactions is superradiance,
 initially theorized by Dicke 
\cite{dicke} and later observed in various systems 
\cite{Srad_atom, Srad_quantumDot, Srad_trapion, Srad_SCqubit}.
These forms of joint quantum phenomena defy
 classical intuition and highlight the
extraordinary nature
of the quantum framework. 

A fundamental aspect of MPMA processes is the subtle 
emergence of a characteristic 
length scale.
 Using 
the two-atom system as an example, let $l_d$
denote the distance between the two atoms.
 It is reasonable to anticipate that the MPMA process 
 is influenced by 
 the value of $l_d$. Specifically, if
$l_d$ exceeds a characteristic length, denoted as 
$l_{mpma}$, the joint excitation process is likely to 
become effectively unsupported.
Since  MPMA processes do not rely on 
 any physical interaction among atoms, $l_{mpma}$
cannot be interpreted as the  range of a specific 
physical force.  This initially makes the determination of
 $l_{mpma}$ seem abstract and elusive.
However, valuable insight emerges naturally from the 
uncertainty principle
 of quantum mechanics. 
The MPMA process can be envisioned as a sequence of 
intermediate virtual transitions,  each involving a temporary 
violation of energy conservation. According to the uncertainty principle, 
the duration of such a virtual transition is inversely 
proportional to the magnitude of the energy deviation. By multiplying 
this timescale by the speed of light, one obtains a spatial range 
within which this virtual transition can collectively contribute 
to the joint excitation.

In the two-photon-two-atom quantum 
process, there are two such virtual transitions. The 
energy mismatch of each transition is:
   $\Delta_\varepsilon = |\hbar \Omega_\mathfrak{L} - \hbar\omega_a|
 =  |\hbar \omega_a - \hbar\omega_b|/2$. Defining
 a length related to this energy mismatch by
 $l_\Delta=  {\hbar c}/{2\Delta_\varepsilon}$, we 
 can approximate the characteristic length as
$l_{mpma}= \alpha l_\Delta$, 
where $\alpha$ is a parameter of order unity or less. 
 This results in a fundamental relationship between $l_{mpma}$
 and the energy mismatch of the process:
\begin{equation}
 l_{mpma}=  \frac{\alpha \hbar c}{2 \Delta_\varepsilon}.
\label{eq:l_DeltaE} 
\end{equation}

We shall consider the two-photon-two-atom process in an 
atomic gas with a large number of atoms. As shown earlier,
the transition rate $W_{2p2m}$ for a single two-atom system,
 is typically small unless the laser
strength reaches an exceptionally high level. This stems from its nature
 as a high-order perturbation process, further 
 diminished by quantum interference effects.
 However, in a gas with a large number of  atoms, the 
 total transition rate for 
 the two-photon-two-atom process rises substantially.

Take an atomic gas with one $A$-species atom and $N_b$ $B$-species 
atoms, confined to a region smaller than $l_{mpma}$. Exposed to the 
same laser field as before, the total rate of the joint two-photon 
absorption increases by a factor of roughly $N_b$, corresponding to 
the number of ways to select one $B$-species atom from the $N_b$ present.

This enhancement follows from Fermi's Golden Rule, which 
implies that the total transition rate is additive over the different final 
quantum states accessible from the same initial state. The system's initial state is
$\Psi_{ini} = |g_a\rangle |g_{b_1}\rangle |g_{b_2}\rangle \ldots |g_{b_{N_b}}\rangle | \scalebox{0.9}{$N_\gamma$} \rangle$,
where $|g_{b_i}\rangle$ ($i = 1, \ldots, N_b$) denotes the ground state 
of the $i$-th $B$-species atom \cite{particleSymmetry}. 
Two-photon excitation involves the $A$-species atom and 
one $B$-species atom, yielding distinct final states. For example, if the $k_1$-th $B$-species atom is excited, the final state becomes:
$\Psi^{ex}_{k_1} = |{e_a}\rangle |g_{b_1}\rangle \ldots |g_{b_{k_1-1}}\rangle |{e_{b_{k_1}}}\rangle |g_{b_{k_1+1}}\rangle \ldots
                       |g_{b_{N_b}}\rangle | \scalebox{0.9}{$N_\gamma-2 $} \rangle$,
with $|e_{b_{k_1}}\rangle$ denoting its excited state.
Summing over all such final states introduces the $N_b$ factor into the total transition rate.

Formally,  let $\Psi(t)$ represent the quantum state at a small time $t$, with $\Psi(0)= \Psi_{ini}$.
 The second-order perturbation expansion yields:
\begin{equation}
\Psi(t) = c_0(t)\Psi_{ini} + \sum\limits_{1\leq k_1 \le N_b} c_{k_1}(t) \Psi^{ex}_{k_1}. 
\end{equation}
Perturbation analysis reveals that  $|c_{1}(t)|^2 \approx |c_{2}(t)|^2 \approx \ldots \approx 
|c_{N_b}(t)|^2 \approx W_{2p2m}\,t$. Consequently, the 
probability of the system remaining unexcited  is $P_{un}(t) = 
|c_0(t)|^2 = 1 - \sum\limits_{1\leq k_1\le N_b} |c_{k_1}|^2  \approx 1 - N_b  W_{2p2m}\, t$. 
The rate of change of $P_{un}(t)$ with respect to $t$ corresponds to the total transition rate.

Extending this to a gas to an atomic gas with
 $N_a$ $A$-species atoms and 
$N_b$ $B$-species atoms, the total transition rate scales by  
$N_a N_b $. 
Since $N_a$ and  $N_b$ can reach values as large as $10^{14}$ or higher, 
simultaneous two-photon absorption  can become 
significant, despite the weakness of a single two-atom transition. 
For an MPMA process involving $m$ atoms—one of species $A$ and $m-1$ of 
species $B$—the enhancement factor scales approximately as $N_a  N_b^{m-1}$.
This amplification could, in principle, make extremely weak atomic 
transitions experimentally accessible, a topic to be explored 
further in a separate study.
 
Consider a homogeneous atomic gas with a linear size exceeding $l_{mpma}$. 
For a given $A$-species atom,  the number of $B$-species atoms participating 
in a two-photon-two-atom transition with it, 
denoted by  $N_{b\mathfrak{o}}$, can be  approximated as  $\rho_b  l_{mpma}^3$,
where $\rho_b$ is the number density of $B$-species atoms.
In this system, the total transition rate of the two-photon-two-atom
process (per $A$-species atom) is proportional to $N_{b\mathfrak{o}} \approx \rho_b
 l_{mpma}^3 $. The characteristic length $l_{mpma}$, which is given by
 $\alpha c  /(\omega_a - \omega_b) $, plays a crucial role in determining the 
 amplification factor of the transition rate.
  Interestingly, the experimental observations of the 
  two-photon-two-atom process \cite{WhiteAtomgasTlBa,NaAtoms} 
  provide valuable comparative insights. 
 These findings suggest  $l_{mpma}$
scales as $ c/(\omega_a - \omega_b)$.

In J. C. White’s study \cite{WhiteAtomgasTlBa}, a mixture of Ba and Tl atoms
 was investigated at
 high temperatures (above 1100$^{\circ}$C) and high atomic
 densities ($10^{16}- 10^{18}$ atoms/cm$^3$). 
 Here, a Ba atom and a Tl atom simultaneously absorbed two laser 
 photons at $2\pi \times 668.1$ THz, despite this frequency being nonresonant with the 
 isolated transitions of Ba ($6s^2 \ ^1S_0 \rightarrow 6p \ ^1P_0$,
  $\omega_{Ba} = 2\pi \times 541.4$ THz) and Tl ($6p \ ^2P_{1/2} \rightarrow 7s \ ^2S_{1/2}$, 
  $\omega_{Tl} = 2\pi \times 793.8$ THz). This process was observed under laser intensities of
 $10^8$–$10^9$ W/cm². The absorption signal was found to be proportional to 
 the square of the laser intensity, confirming a higher-order process 
 involving the simultaneous absorption of two photons. Additionally, it 
 scaled with the product of the Ba and Tl atom densities, providing clear
  evidence of joint excitations between atom pairs.

In contrast, E. Pedrozo-Pe\~nafiel {\it et al.} examined a cold ($100$ $\mu$K), 
dilute Na gas ($10^{12}$ atoms/cm$^3$).  Laser excitation targeted Na dipole
 transitions—$3S_{1/2} \rightarrow 3P_{1/2}$ at $2\pi \times 508.3$ THz 
 and $3S_{1/2} \rightarrow 3P_{3/2}$ at $2\pi \times 508.8$ THz—
 split by a $2\pi \times 0.51$ THz fine-structure gap.
 The system responded to laser light tuned to the average frequency of these transitions.
As in White’s experiments, the excitation signal was proportional to 
the square of the laser intensity.
 Using a relatively low laser intensity (on the order of $100$ mW/cm$^2$)
and a sample of about $10^9$ Na atoms, the two-atom excitation rate was 
observed to be approximately 1\% of the single-atom excitation rate 
under resonant conditions.

The Tl/Ba gas experiment  differs significantly from the Na gas experiment
 due to its much higher atomic density and laser intensity—the latter being at least 
 $10^9$ times greater.
 A thorough comparison should also account for thermal broadening and laser
  linewidth, both of which influence the transition rate.
  In the Tl/Ba gas, thermal broadening exceeded the natural linewidth of dipole 
  transitions by one to two orders of magnitude, while the laser linewidth was
   $10^3$ times broader. Together, these factors suppressed the transition rate by approximately 
   $10^{-4}$ or so. In contrast, these effects were negligible in the cold Na gas.

The transition rate is proportional to the square of the laser intensity and 
either the square of the atomic density (Na gas) or the product of atomic densities (Tl/Ba gas). 
 One key to understanding the disparity 
between the two experiments—particularly the differing laser intensities required
—lies in the characteristic length $l_{mpma}$. This length 
 reflects the energy mismatch of the two-photon-two-atom process, driven 
 by the frequency gap between joint transitions. For the Tl/Ba gas, 
 $l_{mpma} \approx 0.2$ $\mu$m (assuming $\alpha \approx 1$ in 
 Eq.~(\ref{eq:l_DeltaE})),  while for the Na gas, $l_{mpma} \approx 100$ $\mu$m.
 Since the transition rate scales with $l_{mpma}^3$,  this results in a rate difference 
 of approximately $10^8$,  which partially explains why the Tl/Ba experiments
  require much higher laser intensities to observe joint excitations.
Furthermore, the two-photon-two-atom process observed in the dilute 
Na gas suggests that no physical interatomic interaction is 
responsible for the process. In this system, the average interatomic spacing is around
 $1.0 $ $\mu$m, and any residual interatomic interaction is negligible at this distance.

\tikzmath{ \arcR=40; \arcAng=5; \arcH=1.8;  \opacity=0.7; \ballR=0.1cm; 
   \centrX=4.4; \atomShift=0.9; \centrY=-0.9; \bEx=1;\rEx= -0.7;  \Elabsz= 0.85;
   \rEy= -0.03; \bEy=\rEy; \rint= 1.2;\bint=0.8; \bsz=0.5;  \pad=0.05; \seglen= 5pt; \amp=0.6pt;\Esz=0.7; 
\padint= 1.pt; \wsz=0.6; \bEx= \rEx + \Esz+ 0.1; \bsf=0; \wint=0.02; \wx=-1.7; \wyinz=0.1;
\ypanel=3.0;  \levelshift=-0.1; \xpanel=5.0; \wxshift=0.65*\wx; \levelshiftt=-0.1;\wxshiftt=0.65*\wx;}
\begin{figure}
\begin{tikzpicture}[scale=1.0] 

\draw[thick, rcol](\rEx +\levelshift, \rEy)-- +(\Esz, 0);
\node[font = {\tiny},color=rcol] (c) at (\rEx+\Esz*\Elabsz+\levelshift,\rEy+3.5*\pad) {${\varepsilon^a_g}$ };
\draw[thick, rcol](\rEx+\levelshift, \rEy+\rint)--+(\Esz, 0); 
\node[font = {\tiny},color=rcol] (c) at (\rEx+\Esz*\Elabsz+\levelshift,\rEy+\rint+3.5*\pad) {${\varepsilon^a_e}$ };
\draw[thick,rcol](\rEx +\Esz/2+\levelshift, \rEy) circle (0.04);

\draw[dashed, rcol, -stealth](\rEx+\Esz/2+\levelshift, \rEy)-- +(0, \rint);

\draw[thick, bcol](\bEx+\levelshift, \bEy)-- +(\Esz, 0);
\node[font = {\tiny},color=bcol] (c) at (\bEx+\Esz*\Elabsz+\levelshift,\bEy+3.5*\pad) {${\varepsilon^b_g}$ };
\draw[thick, bcol](\bEx+\levelshift, \bEy+\bint)-- +(\Esz, 0);
\node[font = {\tiny},color=bcol] (c) at (\bEx+\Esz*\Elabsz+\levelshift,\bEy+\bint+3.5*\pad) {${\varepsilon^b_e}$ };
\draw[thick,bcol](\bEx +\Esz/2+\levelshift, \bEy) circle (0.04);


\myphotonline.apply(\wx, 0, \wsz, -2*\wint+\wsz, \asz, Laserblue_detune)
\node[font = {\tiny}, color=Laserblue_detune] (c) at (\wx+ 0.5*\wsz , \wyinz) { ...};
\myphotonline.apply(\wx, 2*\wyinz, \wsz, -2*\wint+\wsz, \asz, Laserblue_detune)
\myphotonline.apply(\wx, 3*\wyinz, \wsz, -2*\wint+\wsz, \asz, Laserblue_detune)
\myphotonline.apply_dashed(\wx, 4*\wyinz, \wsz, -2*\wint+\wsz, \asz, Laserblue_detune)
\draw[thick](\wx-0.1, -\wyinz)  -- +(0, 6*\wyinz);
\draw[thick](\wx+ \wsz + 0.1, 5*\wyinz)  -- +(0.1, -3*\wyinz) -- +(0, -6*\wyinz) ;
\myphotonline.apply(\wx, 8*\wyinz, \wsz, -2*\wint+\wsz, \asz, Laserblue_detune)
\draw[dashed, dash pattern={on \padint off \padint on \padint off \padint on \padint}, -stealth]
(\wx + 0.5* \wsz , 4.8*\wyinz )-- +(0, 2.4*\wyinz);

\myphotonline.apply(\wx+\wxshift, 0, \wsz, -2*\wint+\wsz, \asz, redgold)
\node[font = {\tiny}, color=redgold] (c) at (\wx+\wxshift+ 0.5*\wsz , \wyinz) { ...};
\myphotonline.apply(\wx+ \wxshift, 2*\wyinz, \wsz, -2*\wint+\wsz, \asz, redgold)
\myphotonline.apply(\wx+\wxshift, 3*\wyinz, \wsz, -2*\wint+\wsz, \asz, redgold)
\myphotonline.apply(\wx+\wxshift, 4*\wyinz, \wsz, -2*\wint+\wsz, \asz, redgold)
\draw[thick](\wx-0.1+\wxshift, -\wyinz)  -- +(0, 6*\wyinz);
\draw[thick](\wx+\wxshift+ \wsz + 0.1, 5*\wyinz)  -- +(0.1, -3*\wyinz) -- +(0, -6*\wyinz) ;

\draw[thick, rcol](\rEx + \xpanel+\levelshift, \rEy)-- +(\Esz, 0);
\node[font = {\tiny},color=rcol] (c) at (\rEx+\Esz*\Elabsz+ \xpanel+\levelshift,\rEy+3.5*\pad) {${\varepsilon^a_g}$ };
\draw[thick, rcol](\rEx+ \xpanel+\levelshift, \rEy+\rint)--+(\Esz, 0); 
\node[font = {\tiny},color=rcol] (c) at (\rEx+\Esz*\Elabsz+ \xpanel+\levelshift,\rEy+\rint+3.5*\pad) {${\varepsilon^a_e}$ };
\draw[thick,rcol](\rEx +\Esz/2+ \xpanel+\levelshift, \rEy) circle (0.04);


\draw[thick, bcol](\bEx+ \xpanel+\levelshift, \bEy)-- +(\Esz, 0);
\node[font = {\tiny},color=bcol] (c) at (\bEx+\Esz*\Elabsz+ \xpanel+\levelshift,\bEy+3.5*\pad) {${\varepsilon^b_g}$ };
\draw[thick, bcol](\bEx+ \xpanel+\levelshift, \bEy+\bint)-- +(\Esz, 0);
\node[font = {\tiny},color=bcol] (c) at (\bEx+\Esz*\Elabsz+ \xpanel+\levelshift,\bEy+\bint+3.5*\pad) {${\varepsilon^b_e}$ };
\draw[thick,bcol](\bEx +\Esz/2+ \xpanel+\levelshift, \bEy) circle (0.04);

\draw[dashed, bcol, -stealth](\bEx+\Esz/2+ \xpanel+\levelshift, \bEy)-- +(0,\bint);
 
\myphotonline.apply(\wx+ \xpanel, 0, \wsz, -2*\wint+\wsz, \asz, Laserblue_detune)
\node[font = {\tiny}, color=Laserblue_detune] (c) at (\wx+ 0.5*\wsz+ \xpanel , \wyinz) { ...};
\myphotonline.apply(\wx+ \xpanel, 2*\wyinz, \wsz, -2*\wint+\wsz, \asz, Laserblue_detune)
\myphotonline.apply(\wx+ \xpanel, 3*\wyinz, \wsz, -2*\wint+\wsz, \asz, Laserblue_detune)
\myphotonline.apply(\wx+ \xpanel, 4*\wyinz, \wsz, -2*\wint+\wsz, \asz, Laserblue_detune)
\draw[thick](\wx-0.1+ \xpanel, -\wyinz)  -- +(0, 6*\wyinz);
\draw[thick](\wx+ \wsz + 0.1+ \xpanel, 5*\wyinz)  -- +(0.1, -3*\wyinz) -- +(0, -6*\wyinz) ;

\myphotonline.apply(\wx+\wxshift+\xpanel, 0, \wsz, -2*\wint+\wsz, \asz, redgold)
\node[font = {\tiny}, color=redgold] (c) at (\wx+\wxshift+ 0.5*\wsz+ \xpanel , \wyinz) { ...};
\myphotonline.apply(\wx+ \wxshift+\xpanel, 2*\wyinz, \wsz, -2*\wint+\wsz, \asz, redgold)
\myphotonline.apply(\wx+\wxshift+ \xpanel, 3*\wyinz, \wsz, -2*\wint+\wsz, \asz, redgold)
\myphotonline.apply_dashed(\wx+\wxshift+ \xpanel, 4*\wyinz, \wsz, -2*\wint+\wsz, \asz, redgold)
\draw[thick](\wx-0.1+\wxshift+ \xpanel, -\wyinz)  -- +(0, 6*\wyinz);
\draw[thick](\wx+\wxshift+ \wsz + 0.1+ \xpanel, 5*\wyinz)  -- +(0.1, -3*\wyinz) -- +(0, -6*\wyinz) ;
\myphotonline.apply(\wx+\wxshift+ \xpanel, 8*\wyinz, \wsz, -2*\wint+\wsz, \asz, redgold)
\draw[dashed, dash pattern={on \padint off \padint on \padint off \padint on \padint}, -stealth]
(\wx +\wxshift+ 0.5* \wsz+ \xpanel , 4.8*\wyinz )-- +(0, 2.4*\wyinz);

 
 \node[single arrow,draw,  dash pattern={on \padint off \padint on \padint off \padint on \padint}, rotate=90,
      minimum width = 16pt, single arrow head extend= 5pt,
      minimum height=5.0mm] at (\wx + 1.2*\Esz+ 1.*\wsz + 0.0*\xpanel, 0.65* \ypanel  ) {};
      
 \node[single arrow,draw,  dash pattern={on \padint off \padint on \padint off \padint on \padint}, rotate=90,
      minimum width = 16pt, single arrow head extend= 5pt,
      minimum height=5.0mm] at (\wx + 1.2*\Esz+ 1.*\wsz + 1.0*\xpanel, 0.65* \ypanel  ) {};
      
  \node[single arrow,draw,  dash pattern={on \padint off \padint on \padint off \padint on \padint}, rotate= 45,
      minimum width = 16pt, single arrow head extend= 5pt,
      minimum height=5.0mm] at (\wx + 1.2*\Esz+ 1.*\wsz+0.3*\xpanel, 1.75* \ypanel  ) {};

  \node[single arrow,draw,  dash pattern={on \padint off \padint on \padint off \padint on \padint}, rotate= 135,
      minimum width = 16pt, single arrow head extend= 5pt,
      minimum height=5.0mm] at (\wx + 1.2*\Esz+ 1.*\wsz + 0.7* \xpanel , 1.75* \ypanel  ) {};
      
\draw[draw=none, fill=arXivgray, opacity= 0.2,rotate= 0] (\wx-1.5, \rEy+\ypanel-0.5)-- +(0, 1.3*\rint+0.5 ) --
 +(1.3 + 2.5*\Esz+ 2*\wsz,  1.3*\rint+0.5)-- +(1.3+ 2.5*\Esz+ 2*\wsz, 0 )--cycle;

 \draw[draw=none, fill= arXivgray, opacity= 0.2,rotate= 0] (\wx-1.5+\xpanel, \rEy+\ypanel-0.5)-- +(0, 1.3*\rint+0.5 ) --
 +(1.3 + 2.5*\Esz+ 2*\wsz,  1.3*\rint+0.5)-- +(1.3 + 2.5*\Esz+ 2*\wsz, 0 )--cycle;
  
  
 \draw[thick, rcol](\rEx+\levelshiftt, \rEy+\ypanel)--+(\Esz, 0);
\draw[thick, rcol](\rEx+\levelshiftt, \rEy+\rint+\ypanel)--+(\Esz, 0);
\draw[thick,rcol](\rEx +\Esz/2+\levelshiftt, \rEy+\ypanel+\rint) circle (0.04);

 \draw[thick, bcol](\bEx+\levelshiftt, \bEy+\ypanel)--+(\Esz, 0);
\draw[thick, bcol](\bEx+\levelshiftt, \bEy+\bint+\ypanel)-- +(\Esz, 0);
\draw[thick,bcol](\bEx +\Esz/2+\levelshiftt, \bEy+\ypanel) circle (0.04);

\draw[dashed, bcol, -stealth](\bEx+\Esz/2+\levelshiftt, \bEy+\ypanel)--+(0,\bint);

\draw[thick](\wx-0.1, -\wyinz+\ypanel)  -- +(0, 6*\wyinz);
\draw[thick](\wx+ \wsz + 0.1, 5*\wyinz+\ypanel)  -- +(0.1, -3*\wyinz) -- +(0, -6*\wyinz) ;

\myphotonline.apply(\wx, \ypanel, \wsz, -2*\wint+\wsz, \asz, Laserblue_detune)
\node[font = {\tiny},color=Laserblue_detune] (c) at (\wx+ 0.5*\wsz , 1.5*\wyinz+\ypanel) { ...};
\myphotonline.apply(\wx,  3*\wyinz+\ypanel, \wsz, -2*\wint+\wsz, \asz, Laserblue_detune)
\myphotonline.apply(\wx,  4*\wyinz+\ypanel, \wsz, -2*\wint+\wsz, \asz, Laserblue_detune)

\draw[thick](\wx-0.1+\wxshiftt, -\wyinz+\ypanel)  -- +(0, 6*\wyinz);
\draw[thick](\wx+ \wsz + 0.1+\wxshiftt, 5*\wyinz+\ypanel)  -- +(0.1, -3*\wyinz) -- +(0, -6*\wyinz) ;
\draw[dashed, dash pattern={on \padint off \padint on \padint off \padint on \padint}, -stealth](\wx+\wxshiftt + 0.5* \wsz , 4.8*\wyinz+\ypanel )--
+(0, 2.4*\wyinz);
\myphotonline.apply(\wx+\wxshiftt, \ypanel, \wsz, -2*\wint+\wsz, \asz, redgold)
\node[font = {\tiny},color=redgold] (c) at (\wx+ 0.5*\wsz+\wxshiftt,  1.0*\wyinz+\ypanel) { ...};
\myphotonline.apply(\wx+\wxshiftt,  2*\wyinz+\ypanel, \wsz, -2*\wint+\wsz, \asz, redgold)
\myphotonline.apply(\wx+\wxshiftt,  3*\wyinz+\ypanel, \wsz, -2*\wint+\wsz, \asz, redgold)
\myphotonline.apply_dashed(\wx+\wxshiftt,  4*\wyinz+\ypanel, \wsz, -2*\wint+\wsz, \asz, redgold)
\myphotonline.apply(\wx+\wxshiftt,  8*\wyinz+\ypanel, \wsz, -2*\wint+\wsz, \asz, redgold)

 \draw[thick, rcol](\rEx +0.5*\xpanel +\levelshiftt, \rEy+2*\ypanel)--  +(\Esz,0);
\draw[thick, rcol](\rEx +0.5*\xpanel+\levelshiftt, \rEy+\rint+2*\ypanel)-- +(\Esz,0);
\draw[thick,rcol](\rEx +\Esz/2 +0.5*\xpanel+\levelshiftt, \rEy+\rint+ 2*\ypanel) circle (0.04);

  \draw[thick, bcol](\bEx +0.5*\xpanel+\levelshiftt, \bEy+2*\ypanel)-- +(\Esz,0);
\draw[thick, bcol](\bEx+0.5*\xpanel+\levelshiftt, \bEy+\bint+2*\ypanel)--  +(\Esz,0);

\draw[thick,bcol](\bEx +\Esz/2+0.5*\xpanel+\levelshiftt, \bEy+\bint+2*\ypanel) circle (0.04);

\draw[thick](\wx-0.1+0.5*\xpanel, -\wyinz+2*\ypanel)  -- +(0, 6*\wyinz);
\draw[thick](\wx+ \wsz + 0.1+0.5*\xpanel, 5*\wyinz+2*\ypanel)  -- +(0.1, -3*\wyinz) -- +(0, -6*\wyinz) ;

\myphotonline.apply(\wx+0.5*\xpanel, 2*\ypanel, \wsz, -2*\wint+\wsz, \asz, Laserblue_detune)
\node[font = {\tiny},color= Laserblue_detune] (c) at (\wx+0.5*\xpanel+ 0.5*\wsz , \wyinz+2*\ypanel) { ...};
\myphotonline.apply(\wx+0.5*\xpanel,  3*\wyinz+2*\ypanel, \wsz, -2*\wint+\wsz, \asz, Laserblue_detune)
\myphotonline.apply(\wx+0.5*\xpanel,  4*\wyinz+2*\ypanel, \wsz, -2*\wint+\wsz, \asz, Laserblue_detune)

\draw[thick](\wx-0.1+0.5*\xpanel+\wxshiftt, -\wyinz+2*\ypanel)  -- +(0, 6*\wyinz);
\draw[thick](\wx+ \wsz + 0.1+0.5*\xpanel+\wxshiftt, 5*\wyinz+2*\ypanel)  -- +(0.1, -3*\wyinz) -- +(0, -6*\wyinz) ;

\myphotonline.apply(\wx+0.5*\xpanel+\wxshiftt, 2*\ypanel, \wsz, -2*\wint+\wsz, \asz, redgold)
\node[font = {\tiny},color=redgold] (c) at (\wx+0.5*\xpanel+ 0.5*\wsz+\wxshiftt , \wyinz+2*\ypanel) { ...};
\myphotonline.apply(\wx+0.5*\xpanel+\wxshiftt,  3*\wyinz+2*\ypanel, \wsz, -2*\wint+\wsz, \asz, redgold)
\myphotonline.apply(\wx+0.5*\xpanel+\wxshiftt,  4*\wyinz+2*\ypanel, \wsz, -2*\wint+\wsz, \asz, redgold)


    
  
 \draw[thick, rcol](\rEx+\xpanel+\levelshiftt, \rEy+\ypanel)--+(\Esz, 0);
\draw[thick, rcol](\rEx+\xpanel+\levelshiftt, \rEy+\rint+\ypanel)--+(\Esz, 0);
\draw[thick,rcol](\rEx +\Esz/2+\xpanel+\levelshiftt, \rEy+\ypanel) circle (0.04);
\draw[dashed, rcol, -stealth](\rEx+\Esz/2+\xpanel+\levelshiftt, \rEy+\ypanel)--+(0,\rint);
   
 \draw[thick, bcol](\bEx+\xpanel+\levelshiftt, \bEy+\ypanel)--+(\Esz, 0);
\draw[thick, bcol](\bEx+\xpanel+\levelshiftt, \bEy+\bint+\ypanel)-- +(\Esz, 0);
\draw[thick,bcol](\bEx +\Esz/2+\xpanel+\levelshiftt, \bEy+\ypanel+\bint) circle (0.04);


\draw[thick](\wx-0.1+\xpanel, -\wyinz+\ypanel)  -- +(0, 6*\wyinz);
\draw[thick](\wx+ \wsz + 0.1+\xpanel, 5*\wyinz+\ypanel)  -- +(0.1, -3*\wyinz) -- +(0, -6*\wyinz) ;

\draw[dashed, dash pattern={on \padint off \padint on \padint off \padint on \padint}, -stealth]
(\wx + 0.5* \wsz+\xpanel , 4.8*\wyinz+\ypanel )--
+(0, 2.4*\wyinz);
\myphotonline.apply(\wx+\xpanel, \ypanel, \wsz, -2*\wint+\wsz, \asz, Laserblue_detune)
\node[font = {\tiny},color=Laserblue_detune] (c) at (\wx+\xpanel+ 0.5*\wsz , 1.*\wyinz+\ypanel) { ...};
\myphotonline.apply(\wx+\xpanel,  2*\wyinz+\ypanel, \wsz, -2*\wint+\wsz, \asz, Laserblue_detune)
\myphotonline.apply(\wx+\xpanel,  3*\wyinz+\ypanel, \wsz, -2*\wint+\wsz, \asz, Laserblue_detune)
\myphotonline.apply_dashed(\wx+\xpanel,  4*\wyinz+\ypanel, \wsz, -2*\wint+\wsz, \asz, Laserblue_detune)
\myphotonline.apply(\wx+\xpanel,  8*\wyinz+\ypanel, \wsz, -2*\wint+\wsz, \asz, Laserblue_detune)

\draw[thick](\wx-0.1+\xpanel+\wxshiftt, -\wyinz+\ypanel)  -- +(0, 6*\wyinz);
\draw[thick](\wx+ \wsz + 0.1+\xpanel+\wxshiftt, 5*\wyinz+\ypanel)  -- +(0.1, -3*\wyinz) -- +(0, -6*\wyinz) ;

\myphotonline.apply(\wx+\xpanel+\wxshiftt, \ypanel, \wsz, -2*\wint+\wsz, \asz, redgold)
\node[font = {\tiny},color=redgold] (c) at (\wx+\xpanel+ 0.5*\wsz +\wxshiftt, 1.5*\wyinz+\ypanel) { ...};
\myphotonline.apply(\wx+\xpanel+\wxshiftt,  3*\wyinz+\ypanel, \wsz, -2*\wint+\wsz, \asz, redgold)
\myphotonline.apply(\wx+\xpanel+\wxshiftt,  4*\wyinz+\ypanel, \wsz, -2*\wint+\wsz, \asz, redgold)


  

   

\end{tikzpicture}  

\caption{The quantum excitation pathways for the two-laser case. The 
quantum states 
in shaded boxes correspond to 
virtual intermediate states. }  
\label{pathways2lasers}
\end{figure}
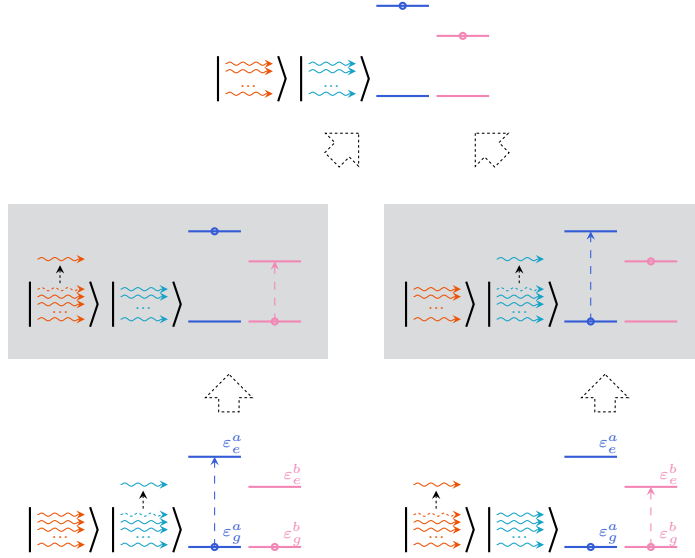

A modified experiment on the two-photon-two-atom phenomenon
can be naturally proposed to investigate the significant role of
 $l_{mpma}$. In this modified setup, a two-laser excitation scheme 
  replaces the single-laser configuration (see Fig. \ref{pathways2lasers} for the
  excitation pathways). Using the Tl/Ba gas 
 system as an example,  
The frequency of one laser (denoted as  $\mathfrak{L_{1}}$) is 
detuned from the Tl transition 
frequency $\omega_{Tl}$  by an amount $\delta_{Tl} = |\omega_{Tl}-\Omega_{\mathfrak{L_{1}}}|$.
Simultaneously, the frequency of the  other laser (denoted as  $\mathfrak{L_{2}}$)
 is detuned from the Ba transition frequency 
by the same  amount in the opposite direction, ensuring the condition  $\Omega_{\mathfrak{L_1}}+ \Omega_{\mathfrak{L_2}} = \omega_{Tl}+ \omega_{Ba}$.   
 In this two-laser scheme, $l_{mpma}= c /2 \delta_{Tl}$.
 By varying $\Omega_{\mathfrak{L_1}}$, $l_{mpma}$
 can be easily tuned,  allowing for making observations of 
 how the 
 detection signal changes with $l_{mpma}$.

 Furthermore, experimental verification of simultaneous 
 three-photon-three-atom processes can be straightforwardly 
 carried out using 
 the two-laser scheme.  In the case of the Tl/Ba gas system,
 the laser frequencies are configured to  satisfy either of the following conditions:
 $ 2\Omega_\mathfrak{L_{1}} + \Omega_\mathfrak{L_{2}}= 2\omega_{Tl} + \omega_{Ba}$, 
 or $ \Omega_\mathfrak{L_{1}} + 2\Omega_\mathfrak{L_{2}}= \omega_{Tl} + 2\omega_{Ba}$.
 In the three-photon-three-atom processes,
The role of $l_{mpma}$   becomes  more pronounced 
compared to the two-photon-two-atom process, as the detection signal scales proportionally 
to $l_{mpma}^6$ rather than $l_{mpma}^3$. 
This heightened sensitivity of the three-photon-three-atom process to variations in 
$l_{mpma}$ provides a more robust platform for studying its influence.

In conclusion, a simultaneous MPMA process in an atomic gas under
 laser fields is examined, revealing several distinctive features,
  including the absence of interatomic interactions and the 
  presence of an intrinsic characteristic length. A proposal for 
  directly demonstrating this characteristic length experimentally 
  is presented. Further studies of this fundamental process 
  will not only enhance our understanding of the interaction 
  between matter and light but also further our exploration of 
  the intriguing quantum world.

\bibliography{mpmabib}

\begin{thebibliography}{23}
\bibitem [{qua()}]{quantum_inference}%
  \BibitemOpen
  \href@noop {} {}\bibinfo {note} {The exact formal cancellation of the
  transition rate of an $m$-atom system at the resonant condition can be
  demonstrated numerically. Moreover, for even $m$, this precise cancellation
  can be justified algebraically in a straightforward manner, as the total
  contribution of all $m!$ excitation pathways exhibits $m!/2$ instances of
  exact pairwise cancellation.}\BibitemShut {Stop}%
\bibitem [{par()}]{particleSymmetry}%
  \BibitemOpen
  \href@noop {} {}\bibinfo {note} {In this analysis, we ignore the atom
  exchange symmetry, as its effect is negligible when the overlaps between the
  individual wavefunctions of $B$-species atoms are minimal, see also
  \cite{dicke}.}\BibitemShut {Stop}%
\end{thebibliography}

\end{document}